\documentstyle{europhys} 

\def\And{{\rm and\ }}

\def\stars{\bigskip\centerline{***}\medskip}

\newif\ifboo \boofalse

\def\Review#1{\boofalse{\it #1},}
\def\Name#1{{\sc #1},}
\def\Vol#1{\ifboo Vol. {\bf #1}\else{\bf #1}\fi}
\def\Year#1{\ifboo #1\else(#1)\fi}
\def\Book#1{\bootrue{\it #1},}
\def\Page#1{\ifboo {\rm p. #1}\else{\rm #1}\fi}

\begin{document}

\euro{43}{2}{226--229}{1998} \Date{15 July 1998} \shorttitle{L. FOSCHINI: 
ELECTROMAGNETIC INTERFERENCE FROM PLASMAS ETC.}

\title{Electromagnetic interference from plasmas\\generated in 
meteoroids impacts} \author{Luigi Foschini\footnote{E-mail: 
L.Foschini@fisbat.bo.cnr.it}} \institute{Istituto FISBAT - 
CNR, Via Gobetti 101, I-40129 Bologna (Italy)}

\rec{3 March 1998}{in final form 25 May 1998}

\pacs{
\Pacs{96}{50Kr}{Meteors and meteoroids} 
\Pacs{52}{50Lp}{Plasma production and heating by shock waves and compression} 
\Pacs{95}{40$+$s}{Artificial Earth Satellites} 
}

\maketitle

\begin{abstract}
	It is shown that the plasma, generated during an impact of a 
	meteoroid with an artificial satellite, can produce 
	electromagnetic radiation below the microwave frequency range.  
	This interference is shown to exceed local noise sources and
        might disturb regular satellite operations.
\end{abstract}

\section{Introduction}
At the end of 1998, the first modules of the \emph{International Space Station} 
will be put in orbit around the Earth and this should open new 
frontiers for life in space.  The intensive use of the space makes 
necessary to know the potential risks.  The threat from meteoroids is 
today well known and several authors have underlined the risks connected 
with the impact on a spacecraft (for a review, see \cite{REV}).  
However, the \emph{Olympus} end-of-life anomaly \cite{CASWELL} and the 
recent work of McDonnell \emph{et al.} \cite{MCDONNELL} put a new 
light on these issues.  The \emph{Olympus} failure is a 
paradigmatic example: in that case, the impact with a Perseid 
meteoroid may have generated electrical failures, leading to a chain 
reaction which culminated with an early end of the mission 
\cite{CASWELL}.  On the other hand, McDonnell \emph{et al.} 
\cite{MCDONNELL} showed that, if the plasma charge and current 
production during an impact are considered, meteoroid streams can be 
very dangerous, even during normal conditions.  It should be noted 
that they considered only damages by direct discharges or current 
injection in circuits (\emph{e.g.} via the umbilical) \cite{MCDONNELL}.

However, there are several other ways by which the plasma could 
interact with the spacecraft electronics.  For example, it is useful 
to recall the work of Cerroni and Martelli \cite{CERRONI}, in which 
they showed that thermal forces in impact-produced plasmas could 
explain the magnetisation observed in the neighbourhood of lunar 
craters.  Even if Cerroni and Martelli studied experimentally 
hypervelocity impacts of aluminium projectiles on basalt targets, it is 
possible to extend their work to general hypervelocity impacts.

Here, we show that a plasma cloud, generated during a 
hypervelocity impact of a meteoroid with an artificial satellite, can 
radiate electromagnetic energy below the microwave frequency range and, 
therefore, may disturb regular satellite operations.

\section{Meteoroids impacts}
It is well known that, during a hypervelocity impact, a fraction of 
the projectile and target materials is evaporated and even ionized 
\cite{FECHTIG}.  A plasma cloud is then created almost instantaneously 
after the impact and expands into the surrounding vacuum.  McDonnell 
\emph{et al.} \cite{MCDONNELL} found an empirical formula for the
evaluation of charge $Q$ produced during a hypervelocity impact.  This 
equation, rearranged in order to emphasize the projectile dimensions 
and density, can be written as follows:

\begin{equation}
	Q\simeq 3.04 \delta^{1.02}r^{3.06}V^{3.48} \ [\mathrm{C}]
	\label{e:chargedim}
\end{equation}

\noindent where $\delta$ is the meteoroid density [kg/m$^{3}$], $r$ 
its radius [m] and $V$ its speed [km/s].  In this paper, we consider, 
as a specific example, the Leonid meteoroid stream, that is the most 
dangerous stream known today owing to its high geocentric speed.  
Typical values of $\delta$ and $V$ for Leonids are, respectively, 
1000~kg/m$^{3}$ and 71~km/s.

In order to calculate the plasma characteristic parameters (Debye 
length $\lambda_{D}$ and plasma frequency $\nu$ \cite{PLASMA}), it 
is necessary to make 
further assumptions.  The plasma cloud generated during an impact has 
been studied theoretically (\emph{e.g.} \cite{SHOCK}) and 
experimentally (\emph{e.g.}, \cite{CERRONI}, \cite{KADONO}, 
\cite{SCHULTZ}).  A projectile in an experimental facility reaches a 
speed up to about 10~km/s, far below meteoroids speeds, but data 
obtained in such experiments allow us to make some extrapolations.

Kadono and Fujiwara \cite{KADONO} recently investigated the expansion 
of the plasma cloud: they used nylon projectiles, with a speed range 
between 3 and 6~km/s, impacting targets made of different materials and 
found that the expansion velocity of the leading edge of the plasma cloud, 
$u_{max}$, is almost constant in time.  The value is about twice that 
of the isothermal sound velocity ($c_{s}$):

\begin{equation}
	u_{max}=\frac{2c_{s}}{\gamma - 1}\sqrt{\frac{\gamma}{3}}= 
	\frac{2\gamma}{\gamma - 1}\sqrt{\frac{RT}{3\mu}}
	\label{e:isosound}
\end{equation}

\noindent where $\gamma=c_{p}/c_{v}$ is the constant ratio of specific 
heats, $R$ is the universal gas constant 
[$R=8.314510$~J$\cdot$mol$^{-1}$K$^{-1}$], $T$ is the gas temperature 
[K] and $\mu$ is the mean atomic weight of the plasma cloud.  For a Leonid 
meteoroid, we can consider an almost complete composition of carbon 
and a plasma temperature of 30,000~K, taking into account that about 
$1\%$ of the kinetic energy is partitioned into ionization 
\cite{FECHTIG}.  For $\gamma$, a value of 1.7 is considered 
\cite{KADONO}.  With these assumptions we obtain an expansion speed 
$u_{max}=12.8\cdot 10^{3}$~m/s.

Moreover, we assume that the plasma cloud is almost hemispherical 
during first 20~$\mu$s and thus has a radius $\rho=0.256$~m.  Then, 
the charge $Q$ generated by an impact of a meteoroid with radius $r$, 
calculated with Equation~(\ref{e:chargedim}), must be distributed 
in a volume about 0.035~m$^{3}$. For the sake of semplicity, we have assumed 
a uniform distribution, even if this is not fully realistic, but it is 
sufficient for our purposes. 
Considering each atom singly ionized, we can 
now calculate the electron volume density $n_{e}$ and, then, the Debye 
length $\lambda_{D}$ and the plasma frequency $\nu$ (see Table~\ref{Tab1}).

\begin{table}
\caption{Examples of charge generation and plasma parameters for 
Leonids. For explanation of symbols, see the text.}
\[
\begin{tabular}{ccccccc}
\hline
\multicolumn{1}{c}{$r$ [m]} & \multicolumn{1}{c}{$Q$ [C]} & 
\multicolumn{1}{c}{$n_{e}$ [m$^{-3}$]} & 
\multicolumn{1}{c}{$\lambda_{D}$ [m]} & \multicolumn{1}{c}{$\nu$ 
[Hz]}\\
\hline
\multicolumn{1}{c}{$10^{-4}$} & \multicolumn{1}{c}{$2.8\cdot 10^{-3}$} 
& \multicolumn{1}{c}{$5.0\cdot 10^{17}$} & 
\multicolumn{1}{c}{$1.7\cdot 10^{-5}$} & \multicolumn{1}{c}{$6.3\cdot 
10^{9}$}\\
\hline
\multicolumn{1}{c}{$10^{-3}$} & \multicolumn{1}{c}{$3.2$} & 
\multicolumn{1}{c}{$5.7\cdot 10^{20}$} & \multicolumn{1}{c}{$5.0\cdot 
10^{-7}$} & \multicolumn{1}{c}{$2.1\cdot 10^{11}$}\\
\hline
\multicolumn{1}{c}{$10^{-2}$} & \multicolumn{1}{c}{$3.7\cdot 10^{3}$} 
& \multicolumn{1}{c}{$6.6\cdot 10^{23}$} & 
\multicolumn{1}{c}{$1.5\cdot 10^{-8}$} & \multicolumn{1}{c}{$7.3\cdot 
10^{12}$}\\
\hline
\end{tabular}
\]
\label{Tab1}
\end{table}

\section{Electromagnetic interferences}
The physical meaning of the plasma characteristic 
parameters ($\lambda_{D}$ and $\nu$) is that electrons can move, with 
respect to ions, a distance $\lambda_{D}$ in a time $\nu^{-1}$, before 
an electric field is developed in order to restore the charge 
neutrality.  Ions and electrons can be seen as electric dipoles, with 
distance $\lambda_{D}$, oscillating with a frequency $\nu$.
If we consider the microwave frequency limit of 
$\nu_{mw}=3\cdot 10^{11}$~Hz, we infer that a Leonid meteoroid with 
a mass up to 8~mg is 
sufficient to generate a plasma cloud with a characteristic frequency 
lower than $\nu_{mw}$.

The average power radiated by a dipole is well known from 
electromagnetic theory \cite{HAUSER}.  We assume that the field 
produced at distances $R>>\lambda_{D}$ (far field condition) by a 
spherical dipole distribution, with $\lambda_{D}$ as radius, is 
equivalent to the field of a point dipole with moment amplitude:

\begin{equation}
	p=\frac{4}{3}\pi e \lambda_{D}^{4} n_{e} \ [\mathrm{C}\cdot 
	\mathrm{m}]
	\label{e:dip}
\end{equation}

\noindent where $e$ is the electron charge [C].  For 
$\nu=\nu_{mw}=3\cdot 10^{11}$~Hz, we obtain $p=1.2\cdot
10^{-23}$~C$\cdot$m. Then, the average power radiated is:

\begin{equation}
	<P>=\frac{p^{2}\omega^{4}}{12\pi\epsilon_{0}c^{3}}\simeq 3\cdot 
	10^{-13} \ [\mathrm{W}]
	\label{e:pow}
\end{equation}

\noindent where $\epsilon_{0}=8.854187817\cdot 10^{-12}$~F/m is 
the vacuum dielectric constant, $c$ is the light speed in vacuum 
[m/s] and $\omega=2\pi\nu$.

This value must be compared with the noise of satellite electronic 
devices.  Below $6\cdot 10^{12}$~Hz, the noise has a flat power 
spectral density of about $4\cdot 10^{-21}$~W/Hz, that is
-204~dB~W/Hz \cite{RADAR}.  If we consider a radar, which has a 
bandwidth of about 60~dB~Hz, and other factors that make worse, 
the mean noise power in a receiver is about -146~dB~W, that is 
$2.5\cdot10^{-15}$~W. Comparing these estimates with (\ref{e:pow}), 
we obtain an interference of at 
least \emph{two order of magnitude greater than the electronic noise}. 
It seems clear therefore that the plasma can generate an 
electromagnetic 
interference that is not negligible and can disturb the regular 
satellite operations. For a specific example, if we consider the 
\emph{International Space Station} (1000~m$^{2}$ area), exposed for 
1~hour to a meteoroid flux like 1966 Leonids, there is $41\%$ of 
impact probability with a meteoroid with mass equal or greater than 
$10^{-8}$~kg \cite{FOSCHINI}. Such an impact flux can produce a 
sequence of interferences which can increase the noise level in 
electronic devices, then disturbing the regular satellite operations.

\section{Conclusions}
After the \emph{Olympus} end-of-life anomaly \cite{CASWELL} and the 
work of McDonnell \emph{et al.} \cite{MCDONNELL}, it seems clear that 
the meteoroids hazard is not restricted to a mechanical damage. 
Here it is suggested a new interference path, that is electromagnetic 
radiation emitted from the impact-produced plasma cloud.  Even if the 
radiated power is not sufficient to destroy anything, it may disturb 
regular satellite operations.  Further investigations should be 
made on specific satellite, because they require detailed information 
about onboard electronics, in order to calculate possible couplings and 
non-linearities.

\stars
Author wishes to thank Paolo Farinella, of Department of Mathematics 
of the University of Pisa, for constructive review.
\newpage

\end{document}